\documentclass[aps,prl,twocolumn,groupedaddress,floatfix,amsmath,showpacs]{revtex4}

\usepackage{graphicx}

\begin{document}
\title{Parametrically-stimulated recovery of a microwave signal using standing spin-wave modes of a magnetic film}
\author{A.V.~Chumak}
 \email{chumak@physik.uni-kl.de}
 \altaffiliation[\\ Also at ]{Department of Radiophysics, Taras Shevchenko National University of Kiev, Kiev, Ukraine.}

\affiliation{Fachbereich Physik and Forschungszentrum OPTIMAS,
Technische Universitat Kaiserslautern, 67663 Kaiserslautern,
Germany}

\author{A.A.~Serga}
\affiliation{Fachbereich Physik and Forschungszentrum OPTIMAS,
Technische Universitat Kaiserslautern, 67663 Kaiserslautern,
Germany}
\author{G.A.~Melkov}
\affiliation{Department of Radiophysics, Taras Shevchenko National University of Kiev, Kiev, Ukraine}
\author{V.~Tiberkevich}
\affiliation{Department of Physics, Oakland University, Rochester,
Michigan 48309, USA}
\author{A.N.~Slavin}
\affiliation{Department of Physics, Oakland University, Rochester,
Michigan 48309, USA}
\author{B.~Hillebrands}
\affiliation{Fachbereich Physik and Forschungszentrum OPTIMAS,
Technische Universitat Kaiserslautern, 67663 Kaiserslautern,
Germany}

\date{\today}

\begin{abstract}
The phenomenon of storage and parametrically-stimulated recovery of a microwave signal in a ferrite film
has been studied both experimentally and theoretically. The microwave signal is stored in the form of
standing spin-wave modes existing in the film due to its finite thickness. Signal recovery is performed
by means of frequency-selective amplification of one of these standing modes by double-frequency
parametric pumping process. The time of recovery, as well as the duration and magnitude of the recovered
signal, depend on the timing and amplitudes of both the input and pumping pulses. A mean-field theory of
the recovery process based on the competitive interaction of the signal-induced standing spin-wave mode
and thermal magnons with the parametric pumping field is developed and compared to the experimental
data.

\end{abstract}


\maketitle%

\section{Introduction}

The problem of microwave information storage and processing  using
elementary excitations of matter has been intensively studied both
theoretically and experimentally. For a long time the search
concentrated on different types of echo-based phenomena involving
phase conjugation techniques \cite{Echo_Review, WFR_1}.

Several years ago, a new method of signal restoration was proposed and tested in the experiments with
dipolar spin waves scattered on random impurities and defects of a ferrimagnetic medium
\cite{Reversal_of_Momentum_Relaxation}. In the framework of this method, to achieve the signal
restoration, a frequency-selective parametric amplification of a narrow band of scattered spin waves
(having frequencies close to the frequency of the input signal) was used. As a result of this selective
amplification, a uniform distribution of the secondary (scattered) spin waves in the phase space of the
system was distorted, and a macroscopic noise-like signal was registered at the output
\cite{Methods_of_relaxation_reversal}. The noise-like character of the restored signal is caused by the
fact that this signal is formed by many individual scattered spin waves, having close, but arbitrarily
shifted phases.

\begin{figure}[t]
\begin{center}
\scalebox{1}{\includegraphics[width=7.5 cm,clip]{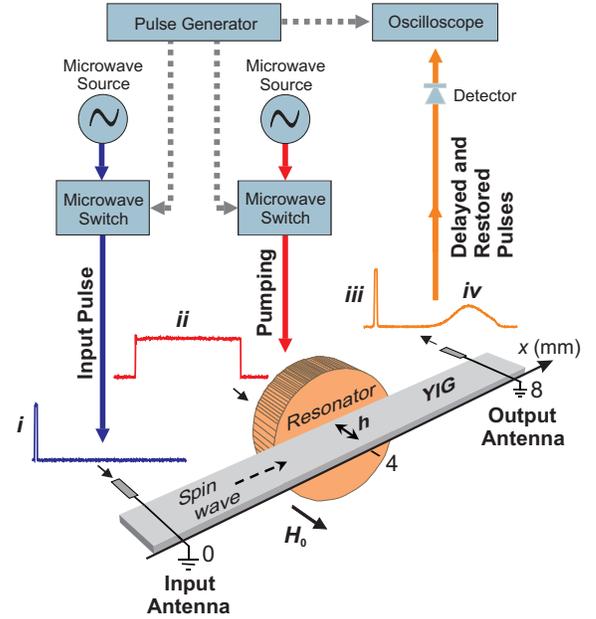}}
\end{center}
\vspace*{-0.4cm}\caption{(color online) Experimental setup and typical waveforms: ``i'' -- input pulse;
``ii'' -- pumping pulse; ``iii'' -- delayed running spin-wave pulse; ``iv'' -- restored signal.}
\label{Setup}
\end{figure}

Recently \cite{PRL_restoration}, we have reported experimental results on storage and recovery of a
microwave signal using a \textit{single}  standing spin-wave mode belonging to the discrete spin-wave
spectrum, which is caused by the spatial confinement of the magnon gas in a thin yttrium-iron-garnet
(YIG) ferrite film.  The storage effect was realized through the conversion of the input microwave
signal into a propagating  magnetostatic wave (or dipolar spin wave), and, then, into an
exchange-dominated standing spin-wave modes (or thickness modes) of the film. The recovery of the signal
was performed by means of frequency-selective parametric amplification, but, in contrast with
\cite{Reversal_of_Momentum_Relaxation}, the restored signal was formed by a single standing spin-wave
mode of the film, and had a practically noiseless character. The mechanism of such a restoration is
highly non-trivial. Therefore, in our first report \cite{PRL_restoration} we used an approximate
empirical theoretical model to explain this rather complicated restoration process.

In our present paper we give a detailed theoretical explanation of the  restoration effect. This
explanation is based on the general theory of parametric interaction of spin waves (so-called
``S-theory'') \cite{S-theory_Lvov_book} and takes into account interactions between different groups of
spin waves existing in a ferrite film. Using the developed theory we calculate parameters of the
restored pulse (power, duration, and delay in respect to the input pulse) as functions of the power of
the input and pumping pulses, and compared these calculated parameters with experimental data.

\section{Experiment}

For the reason of completeness we report here again the experimental findings presented in
\cite{PRL_restoration}, amended by additional experimental investigations. The experimental setup is
shown in Fig.~\ref{Setup}. The input electromagnetic microwave pulse is converted by the input
microstrip transducer into dipolar spin waves, that propagate in a long and narrow ($30 \times
1$\,mm$^2$), 5\,$\mu$m thick yttrium iron garnet (YIG) film waveguide (saturation magnetization $4 \pi
M_\mathrm{s}=1750$\,G, exchange constant $D=5.4 \cdot 10^{-9}$\,Oe$\cdot$cm$^2$). The other transducer,
used to receive the output microwave signals, is placed at a distance of $l=8$\,mm from the input one. A
bias magnetic field of $H_0=1706$\,Oe is applied in the plane of the YIG film waveguide, along its width
and perpendicular to the direction of the spin wave propagation.

The input rectangular electromagnetic pulses having a duration of 100\,ns, carrier frequency of
$f_\mathrm{in}=7.040$\,GHz, and varying power of 0.1\,$\mathrm{\mu W} <P_\mathrm{in}<6$\,mW were
supplied to the input transducer. These input pulses excite wave packets of magnetostatic surface waves
(MSSW) (or Damon-Eshbach magnetostatic waves \cite{Magnetostatic_Modes_of_Ferromagnetic_Slab}), having a
carrier wave number $k_\mathrm{in} \simeq 100$\,cm$^{-1}$. The group velocity $v_\mathrm{g}$ of the
excited wave packet is about 2.3\,cm$/ \mu$s. This group velocity determines the time delay of the
propagating wave packet between the input and output transducers of about 350\,ns. The spin-wave packet
received by the output transducer is, again, converted into an electromagnetic pulse. After
amplification and detection, the output signal is observed with an oscilloscope.

As it was pointed out in \cite{PRL_restoration}, the propagating MSSW excited in the ferrite film
standing (thickness) spin-wave modes that continued to exist in the film long after the propagating MSSW
signal reach the output transducer. To recover the microwave signal stored in these standing spin-wave
modes it is necessary to apply to the film an external pumping microwave field with the frequency that
is approximately two times larger than the carrier frequency of the stored microwave signal. To supply
this double-frequency pumping pulse an open dielectric resonator is placed in the middle of the YIG
waveguide. The resonator is excited by an external microwave source and produces a pumping magnetic
field $\vec{h}_\mathrm{p}$ that is parallel to the static bias magnetic field $\vec{H}_0$, see
Fig.~\ref{Setup}. The resonance frequency of the pumping dielectric resonator $f_\mathrm{r}=14.078$\,GHz
was chosen to be close to twice the carrier frequency of the input microwave pulse. Thus, the conditions
for the process of parallel parametric pumping \cite{Magnetization_Oscillations_and_Waves} are fulfilled
in our experiment. Under these conditions the energy of the pumping field is most effectively
transferred to the magnetic oscillations and waves that have  the frequency that is exactly half of
frequency of the pumping signal.

The experiment starts at the initial moment of time $t=0$ when an external microwave pulse is supplied
to the input transducer (see waveform ``i'' in Fig.~\ref{Setup}). The MSSW packet, excited at the input
transducer by this signal, propagates to the output transducer and creates there a delayed output
microwave signal of similar duration (see waveform ``iii'' in Fig.~\ref{Setup}). The delay time
$t_\mathrm{prop}$ of this output pulse is determined by the MSSW group velocity and the distance between
the transducers. Then, at the time $t>t_\mathrm{prop}$, a relatively long (see waveform ``ii'' in
Fig.~\ref{Setup}) and powerful pumping pulse with a carrier frequency approximately twice as large as
the carrier frequency of the input pulse is supplied to the pumping resonator. Then, during the action
of the pumping pulse the restored signal appears after a certain delay at the output transducer (see
waveform ``iv'' in Fig.~\ref{Setup}).

\begin{figure}
\begin{center}
\scalebox{1}{\includegraphics[width=8.0 cm,clip]{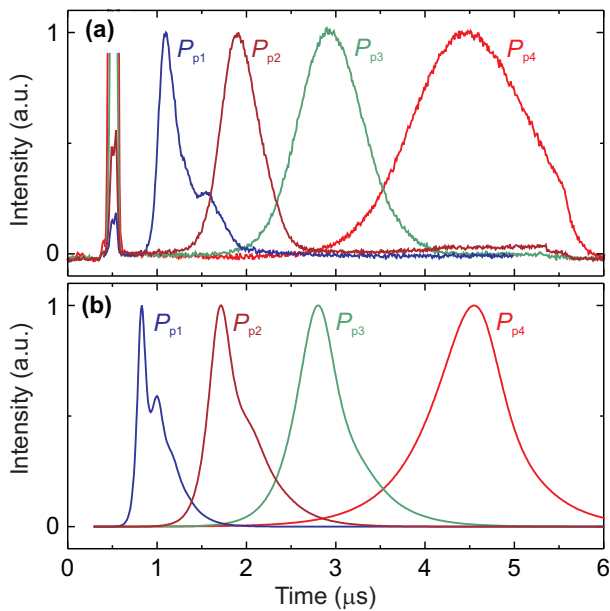}}
\end{center}
\vspace*{-0.7cm}\caption{(color online) Experimental (a) and
calculated (b) waveforms of the delayed and recovered pulses for
different values of the pumping power $P_\mathrm{p}$:
$P_\mathrm{p1}=3.67$\,W, $P_\mathrm{p2}=1.28$\,W,
$P_\mathrm{p3}=0.52$\,W, $P_\mathrm{p4}=0.34$\,W. Waveforms are
normalized by the maximum intensity of each of the recovered
pulses.} \label{Waveforms}
\end{figure}

It is necessary to stress here, that the parametrically recovered signal (waveform ``iv'' in
Fig.~\ref{Setup}) originates from the input microwave signal (waveform ``i'' in Fig.~\ref{Setup}, and
never appears without the previous application of the input signal. At the same time, the main
characteristics of this restored pulse (such as peak power, duration, and delay time) are not directly
related to similar parameters of the input pulse, and are mainly determined by the process of parametric
interaction of the \textit{trail} of the input signal (in the form of standing spin-wave modes) with the
pulsed parametric pumping.

Typical experimental oscillograms demonstrating the normalized waveforms of the restored output pulse
measured for different values of the pumping power are presented in Fig.~\ref{Waveforms}(a). It is
evident from Fig.~\ref{Waveforms}(a) that the increase of the pumping power $P_\mathrm{p}$ leads to a
significant  variation of the restored pulse parameters: decrease of the recovery time $t_\mathrm{r}$
and decrease of the restored pulse duration $\Delta t_\mathrm{r}$. The experiment shows (although it is
not seen in the normalized graphs of Fig.~\ref{Waveforms}(a)) that the peak power $P_\mathrm{r}$ of the
restored pulse increases with the increase of the pumping power $P_\mathrm{p}$.

Figure~\ref{Waveforms}(b) shows the profiles of the output restored pulse calculated using the
theoretical model presented below. It is clear from the comparison of  the experimental and theoretical
waveforms presented in Fig.~\ref{Waveforms}(a),(b) that this model gives a good quantitative description
of the experimental data. In the following section we present a detailed description of this theoretical
model.

\begin{figure}
\begin{center}
\scalebox{1}{\includegraphics[width=8.0cm,clip]{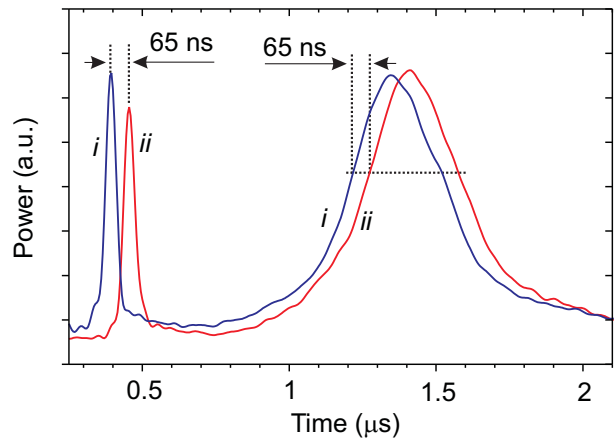}}
\end{center}
\vspace*{-0.5cm}\caption{(color online) Waveforms of the delayed and restored pulses received on the
output antennae placed 6.5~mm (waveform ``i'') and 8~mm (waveform ``ii'') apart from the input antenna.}
\label{two_antennae}
\end{figure}

Looking at Fig.~\ref{Waveforms} one can notice that the recovery time of the restored pulse is much
larger than the time of the spin wave propagation between the input and output transducers. This means
that very slow spin-wave modes participate in the process of signal storage and restoration. In order to
better understand the signal storage mechanism an additional experiment was performed. In this
experiment an additional (second) receiving transducer was placed at a distance of 1.5~mm from the first
(main) one. The time profiles of the spin-wave signal received at the main and additional transducers
are presented in Fig.~\ref{two_antennae}. These profiles consist of two peaks: the first narrow peak,
corresponding to the propagating MSSW packet excited by the input microwave pulse, and the second
broader peak, corresponding to the stored and, then, parametrically recovered spin-wave packet.

One can see that the time interval between the appearance of the fronts of these two peaks at a
particular transducer is the same. This result means that the group velocities of the two delayed
spin-wave packets, which are detected by the two output transducers are the same within the error
margins. Thus, the excitation of the second (restored) peak at the  output transducer is performed by
the fast propagating MSSW mode, the same spin-wave mode that was excited initially by the input
microwave pulse and which created the first narrow delayed peak at the output transducer.

The experimental result shown in Fig.~\ref{two_antennae}, indicates that, apart from the propagating
MSSW, a different spin-wave mode having a negligible group velocity (or, in other words, a standing
spin-wave mode) must take part in the signal storage and recovery process. This standing spin-wave mode
stores the information about the input microwave signal for a couple of microseconds and, then, is
amplified by a parametric pumping pulse supplied to the open dielectric resonator. It is then converted
into a propagating MSSW packet. Obviously, the parameter that is the most important one for this
recovery process is the amplitude of the standing spin-wave mode in the immediate vicinity of the
pumping resonator, as the pumping field of the resonator can only effectively interact with spin waves
located near the resonator.

\section{Qualitative model}

\subsection{Dipole-exchange spin-wave spectrum of a ferrite film}

\begin{figure}
\begin{center}
\scalebox{1}{\includegraphics[width=8.0 cm,clip]{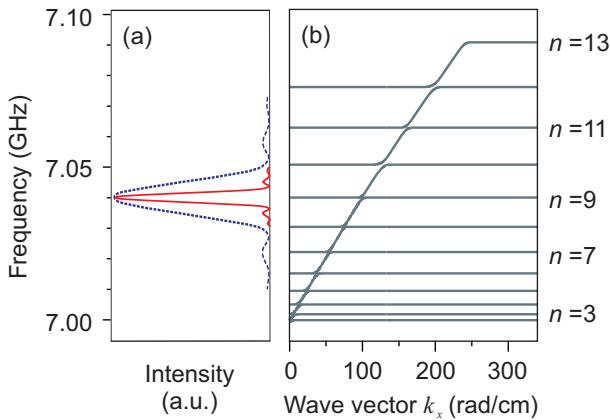}}
\end{center}
\vspace*{-0.5cm}\caption{(color online) (a) Frequency spectra of the
input pulse (blue, dotted line) and pumping pulse (transferred to
the half of the carrier pumping frequency) (red, solid line); (b)
Calculated dipole-exchange spectrum of traveling spin waves in the
experimental YIG film ($n$ is a number of a corresponding thickness
mode).} \label{Spectrum}
\end{figure}

To understand the mechanism governing the observed storage-and-recovery effect of a microwave signal in
a ferrite film of a finite thickness $L$ let us consider the dipole-exchange spectrum of such a film in
the case of the MSSW geometry, when the spin-wave carrier wave number $k_x$ is in the film plane and
perpendicular to the bias magnetic field $H$. A calculated spin-wave spectrum for this case is shown in
Fig.~\ref{Spectrum}(b). It consists of a dipole-dominated (Damon-Eshbach-like
\cite{Magnetostatic_Modes_of_Ferromagnetic_Slab}) spin-wave mode, having the largest group velocity, and
a series of exchange-dominated thickness spin-wave modes of the film having very low group velocities
(see \cite{Kalinikos86} for details). These exchange-dominated modes are close to the standing thickness
modes of a spin-wave resonance of the film, and in the simple case of unpinned surface spins these modes
have discrete values of the perpendicular (to the film plane) wave vector defined as  $ k_{\perp} = \pi
n/L$, where $n=1,2,3...$. Near the crossing points of the lowest ($n=0$) and higher-order spin-wave
modes the dipolar hybridization of the spin-wave spectrum takes place, and the so-called ``dipolar
gaps'' in the spin-wave spectrum of the film are formed \cite{Kalinikos86}.

The qualitative picture of the storage and recovery of the microwave signal in ferrite film looks as
follows. A relatively short (duration 100\,ns) input microwave pulse supplied to the input transducer
excites a packet of propagating MSSW in the ferrite film waveguide. Under the conditions of our
experiment the frequency separation between the discrete quasi-standing spin-wave modes of the film
spectrum is around 10-20\,MHz, depending on the mode number $n$. At the same time, the width of the
frequency spectrum of a relatively short (duration 100\,ns) input microwave pulse, which excites a
propagating MSSW packet in the film, is several times larger. Therefore, this packet can simultaneously
excite \textit{several} quasi-standing spin-wave modes in the film. Due to their extremely low group
velocity these quasi-standing spin-wave modes do not propagate away from the point where they are
excited. Instead, they form a ``trail'' along the path of the propagating MSSW packet, and  this
``trail'' exists for over a microsecond after the MSSW packet itself is gone from the film.

It is worth noting, that the ``trail'' of quasi-standing spin-wave
modes  is excited mainly in the spectral regions of frequency
hybridization (``dipolar gaps'') in the dipole-exchange spectrum of
the film (see Fig.~\ref{Spectrum}) \cite{Kalinikos86}. The amplitude
of this "trail" decays exponentially with time due to the natural
magnetic dissipation in the film.

To observe the restoration of the microwave signal from the ``trail'' it is necessary to supply the
pumping pulse at the time, when the ``trai'' has not yet decreased to the thermal level. When the
pumping is applied, the amplitudes of the quasi-standing spin-wave modes forming a ``trail'' start to
increase due to the parametric amplification. If the pumping pulse is long enough, so that the pumping
has a relatively narrow frequency spectrum and is, therefore, frequency-selective, only \textit{one}
quasi-standing spin-wave mode is amplified parametrically.

The increase of the amplitude of this quasi-standing mode is, eventually, limited by nonlinear spin-wave
interaction processes that will be discussed in detail below. One of the most important nonlinear
processes of this kind is the interaction of the parametrically amplified quasi-standing dipole-exchange
spin-wave mode with the packet of exchange-dominated spin waves, that are excited by the parallel
pumping from the thermal level \cite{JETP-99}.

At the same time, the back-conversion of the parametrically amplified quasi-standing  spin-wave mode
into a propagating MSSW packet takes place in the frequency interval near the "dipole gap" that is
resonant with the pumping carrier frequency. This process results in the formation of the restored
microwave signal at the output transducer (see the waveform ``iv'' in Fig.~\ref{Setup}).

We would like to emphasize one more time, that the restored delayed pulse (waveform ``iv'' in
Fig.~\ref{Setup}) is \textit{not} a product of a direct parametric amplification of the propagating MSSW
packet, since it can be observed even if the pumping pulse is supplied {\it after} the input MSSW has
passed the pumping resonator. Moreover, at a given carrier frequency of the input microwave pulse, the
recovered pulse is observed only if the half-frequency of the microwave pumping lies inside a narrow
frequency interval close to the position of one of the the dipole gaps in the frequency spectrum of the
magnetic film (see e.g. Chapter~7 in \cite{Magnetization_Oscillations_and_Waves} and
\cite{Kalinikos86}). The observation of the restored spin-wave packets only in these narrow frequency
intervals suggests that the process of storage and restoration of the initial microwave signal is caused
by the excitation of the {\it discrete} thickness-related quasi-standing spin-wave modes of the magnetic
film.

\subsection{Storage and restoration of a signal as a multi-step process involving  several groups of spin waves}

In order to explain the above described signal restoration effect a model of interaction of two magnon
groups with parametrical pumping was proposed in \cite{PRL_restoration}. In this model we analyze the
interaction between the electromagnetic parametric pumping with effective amplitude $V h_\mathrm{p}$ and
frequency $\omega_\mathrm{p}$ (where $h_\mathrm{p}$ is a variable pumping magnetic field, $V$ is the
parametrical coupling coefficient), standing spin-wave mode with effective amplitude characterized by
the magnon number $N_\mathrm{s}$ and frequency $\omega_\mathrm{p}/2$, and, the so-called, ``dominating''
spin-wave group with effective amplitude characterized by the magnon number $N_\mathrm{\kappa}$ and
frequency $\omega_\mathrm{p}/2$. In this model the condition for the energy conservation
\cite{S-theory_Lvov_book} is always fulfilled automatically, because the frequency of all the waves
taking part in the effective parametric interaction with pumping is twice smaller than the frequency of
pumping. We call the group of spin waves that is excited by parametric pumping from the thermal level
``dominating'' because this group of waves has the smallest relaxation parameter. Therefore, if the
pumping acts for a sufficiently long time, so that the system reaches a saturated stationary regime,
this dominating group suppresses all the other spin waves taking part in the process of parametric
interaction. The initial amplitude of the dominating group, characterized by the magnon number
$N_\mathrm{\kappa0}$, is determined by the thermal noise level, while the initial amplitude of the
standing spin-wave mode, characterized by the magnon number $N_\mathrm{s0}$, is determined by the
amplitude of the applied input signal, relaxation parameter of this mode, and the time delay between the
signal and pumping pulses.

The process of signal restoration involves competition of two wave groups: the signal-induced standing
wave group and the noise-induced dominating wave group, while both these groups are parametrically
amplified by pumping. The relative efficiency of this parametric amplification is determined by the
relaxation parameters of the wave groups, and, as it was mentioned above, the dominating spin-wave group
has the smallest relaxation parameter $\Gamma_\mathrm{\kappa}$. Thus, the amplification of this group is
higher in comparison to the amplification of all the other spin-waves groups, including the standing
wave group. The rapid increase of the amplitude of the dominating group of spin waves will lead to its
interference with the standing spin-wave group, and, eventually, to the decrease of parametric
amplification and saturation of dominating group amplitude in the stationary regime. Simultaneously, the
amplitude of the standing spin-wave mode, which is competing with the dominating group for the energy
from the pumping, will decrease, because in this  nonlinear competition process the mode with larger
amplitude (i.e. the dominating mode) will get a proportionally larger share of the pumping energy
\cite{S-theory_Lvov_book}.

\begin{figure}[t]
\begin{center}
\scalebox{1}{\includegraphics[width=8 cm,clip]{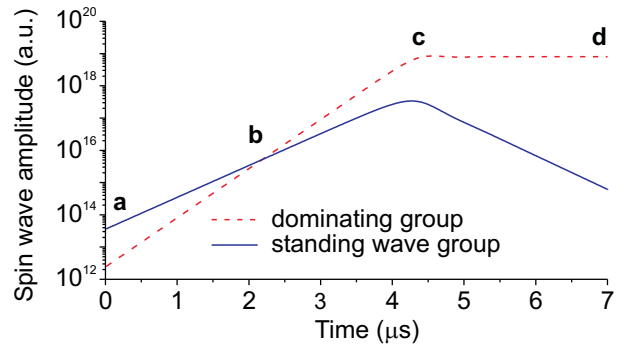}}
\end{center}
\vspace*{-0.5cm}\caption{(color online) Qualitative picture of the temporal evolution of the amplitudes
of standing (solid line) and dominating (dashed line) spin-wave groups interacting with the
constant-amplitude parametric pumping that was switched on at $t=0$. Point ``\textbf{a}'' -- start of
parametrical amplification; ``\textbf{b}'' -- point where the amplitude of the dominating spin-wave
group becomes larger than the amplitude of the standing wave group; ``\textbf{c}'' -- saturation point
where the parametrical amplification stops; ``\textbf{d}'' -- stationary regime where the amplitude of
the dominating group is constant and the amplitude of the standing wave group vanishes.} \label{Scheme}
\end{figure}

A qualitative sketch of the temporal evolution of the amplitudes of standing (solid line) and dominating
(dashed line) wave groups interacting with constant-amplitude parametric pumping is presented in Fig.~
\ref{Scheme}. At the initial time point, when the pumping is switched on, (point ``\textbf{a}'' in the
figure) the signal-induced standing spin-wave mode (solid line) has the amplitude that is larger than
the amplitude of the noise-induced dominating spin-wave group (dashed line). In the region between the
points ``\textbf{a}'' and ``\textbf{c}'' the amplitudes of both wave groups exponentially increase due
to the parametric amplification: the amplitude of the dominating group increases with time $t$ as
$\exp[( h_\mathrm{p} V-\Gamma_\mathrm{\kappa}) t]$, while the amplitude of the signal-induced standing
wave group increases as $\exp[( h_\mathrm{p} V-\Gamma_\mathrm{s} ) t]$ (where $\Gamma_\mathrm{\kappa} <
\Gamma_\mathrm{s}$ are the relaxation parameters of the dominating and standing wave groups
correspondingly and $V$ is the coefficient of parametric coupling between the wave group and the pumping
field $h_\mathrm{p}$).

It is clear that the amplification of the dominating group is larger, because of the smaller relaxation
parameter for this wave group, and beyond the point ``\textbf{b}'' in Fig.~\ref{Scheme} the amplitude of
the dominating group overtakes the amplitude of the standing mode. Then, at the point \textbf{c} the
amplitude of the dominating wave group reaches a certain critical value after which it starts to
renormalize the effective pumping, which leads to saturation and stop of parametric amplification. The
effective pumping amplitude $ h_\mathrm{p}^{\mathrm{eff}} V$ at the saturation point ``\textbf{c}''
becomes equal to the relaxation parameter of the dominating waves with the smallest relaxation
\cite{S-theory_Lvov_book}.

For the conditions of our experiment this means that $h_\mathrm{p}^{\mathrm{eff}} V =
\Gamma_\mathrm{\kappa}$ in the stationary regime (region between ``\textbf{c}'' and ``\textbf{d}'').
Since $ h_\mathrm{p}^{\mathrm{eff}} V < \Gamma_\mathrm{s}$ the amplitude of the signal-induced standing
wave group start to decrease beyond the point \textbf{c} with the factor
$\exp[(h_\mathrm{p}^{\mathrm{eff}} V-\Gamma_\mathrm{s} ) t] = \exp[( \Gamma_\mathrm{\kappa} -
\Gamma_\mathrm{s} ) t]$, as it is shown in the figure. Thus, the restored signal reaches its maximum
amplitude at the time point ``\textbf{c}''.

The mechanism of saturation of parametric amplification leading to the decrease of the restored signal
amplitude can be explained in the framework of the general nonlinear theory of parametric wave
interaction \cite{S-theory_Lvov_book, Magnetization_Oscillations_and_Waves}. According to this theory
there are three main processes which can limit the parametric amplification: nonlinear frequency shift,
nonlinear dissipation, and the so-called phase mechanism of amplification limitation. It was, however,
established in \cite{S-theory_Lvov_book} that it is the phase mechanism that, for the most part, limits
the parametric amplification in the process of parallel pumping used in our current experiment. This
phase mechanism is based on the idea that the nonlinear interaction between pairs of spin waves
parametrically excited by pumping creates a nonlinear shift of their phase, which leads to the decrease
of the efficiency of the pairs interaction with pumping \cite{S-theory_Lvov_book}.

It is well-known, that in order to fulfill the conditions of parametric amplification of a spin wave
(having wave vector $k$ and frequency $\omega_k$) by an external electromagnetic pumping it is necessary
to fulfil the conservation laws for both energy and wave vector \cite{Parametrical_Ampl}. If the pumping
is quasi-uniform, i.e. the pumping wave vector is small $k_\mathrm{p}\approx0$, the interaction of the
signal wave  $c_k$ with pumping leads to the appearance of the ``idle'' wave $c_{-k}$ which has the wave
vector $-k$ and forms a pair $(k,-k)$ with the signal wave. The sum of phases in the pair is fully
determined by the phase of the pumping according to the equation \cite{S-theory_Lvov_book}:
\begin{equation}\label{phases}
\varphi_k + \varphi_{-k} = \varphi_\mathrm{p} + \pi/2,
\end{equation}
where $\varphi_k$ is the signal wave phase, $\varphi_{-k}$ is the idle wave phase and
$\varphi_\mathrm{p}$ is the pumping phase (including the phase of the coupling coefficient between
pumping and spin waves).

When the spin-wave amplitudes are small Eq.~(\ref{phases}) can  be easily satisfied as the ``idle'' wave
with the proper phase can be chosen from the multitude of thermally excites spin waves. With the
increase of the spin-wave amplitudes the pairs of the parametrically excited spin waves start to
interact with each other through the four-wave process of pair interaction. As a result of this
interaction process the sum of phases $\psi_k =  \varphi_k + \varphi_{-k}$ starts to change, the
condition (\ref{phases}) breaks, and the efficiency of the pumping-induced parametric amplification of
the signal waves is drastically reduced \cite{S-theory_Lvov_book}.

In our first paper \cite{PRL_restoration} describing the signal restoration process we obtained analytic
expressions for the time of appearance and the amplitude of the parametrically restored pulse, but did
not analyze the nonlinear processes leading to the limitation of parametric amplification and,
therefore, to the finite duration of the restored pulse. Below, we present a detailed theoretical model
of the signal restoration where all the relevant nonlinear processes of wave interaction are taken into
account.

\section{Theoretical model}

In order to explain the above described effect of microwave signal restoration we use the general theory
of parametric interaction of spin waves (so-called ``S-theory'') \cite{S-theory_Lvov_book}. The equation
for the amplitude of a spin wave interacting with the parametric pumping can be written in the form (see
also \cite{Magnetization_Oscillations_and_Waves}):
\begin{equation}\label{base1}
\left[\frac{d}{dt}+\Gamma - i
(\tilde{\omega}_k-\omega_\mathrm{p}/2)\right] c_k - i P_k c_{-k}^* =
0,
\end{equation}
where $c_k$ and $c_{-k}^{*}$ are the amplitudes of signal and idle waves of frequency $\omega_k$ and
wave number $k$, $\Gamma$ is the spin-wave relaxation parameter, and $\omega_\mathrm{p}=2 \pi
f_\mathrm{p}$ is the pumping frequency.
\begin{equation}\label{omega_eff}
\tilde{\omega}_k = \omega_k + 2 \sum_{k_1}{T_{k k_1} |c_{k_1}|^2}
\end{equation}
is a spin-wave frequency with account of nonlinear frequency shift, $T_{k k_1}$ is the corresponding
nonlinear parameter,
\begin{equation}\label{Pk_eff1}
P_k = h_\mathrm{p} V + \sum_{k_1}{S_{k k_1} c_{k_1} c_{-k_1}}
\end{equation}
is the effective internal amplitude of parametric pumping, and $S_{k k_1}$ is the nonlinear coefficient
describing four-wave interaction of spin-wave pairs. The analogous equation for the amplitude of the
idle wave $c_{-k}^{*}$ is omitted for brevity.

The spin-wave formalism is substantially simplified if, instead of the individual complex amplitudes of
the signal and idle waves, we introduce new variables characterizing combined amplitudes (or magnon
densities per unit volume for a given wave number $k$) and phases of the spin-wave pairs, following
\cite{S-theory_Lvov_book}:
\begin{equation}\label{pair n}
n_k = M_0/ 2(\gamma \hbar/) c_{k} c_{-k}e^{-i \psi_k}
\end{equation}
\begin{equation}\label{pair psi}
\psi_k = \varphi_k + \varphi_{-k},
\end{equation}
where $\gamma = 2.8$\,MHz/Oe is the gyromagnetic ratio, $\hbar$ is
the Planck constant, and  $M_0$ is the saturation magnetization.

In these ``pair'' variables the equations for the amplitudes of parametrically interacting spin waves
can be written in a simple form \cite{S-theory_Lvov_book}:
 equation (\ref{base1}) can be presented in a form:
\begin{eqnarray}\label{base2}
&&\frac{1}{2}\frac{dn_k}{dt} = n_k
[-\Gamma_k  + \mathrm{Im}(P_k^{*} e^{i \psi_k})] \nonumber\\
&&\frac{1}{2}\frac{d \psi_k}{dt} =
\tilde{\omega}_k-\omega_\mathrm{p}/2 + \mathrm{Re}(P_k^{*} e^{i
\psi_k}).
\end{eqnarray}

In our model we analyze the existence of two magnon groups (or two effective magnon pairs): the
dominating group with magnon density $n_\mathrm{\kappa}$ (note, that index $\mathrm{\kappa}$ is used to
mark this dominating spin-wave group) and the signal-induced quasi-standing group of spin waves with
magnon density $n_s$ \cite{PRL_restoration}.

Below, we shall assume for simplicity that the coefficient of four-wave  pair interaction for all the
wave groups has approximately the same value  $S_{k k_1} = S_{0 0}$,  and  that the effective pair phase
is approximately equal for both spin-wave groups (dominating and standing) involved in the parametric
interaction with pumping. Under these assumptions we can substantially simplify the expression
(\ref{Pk_eff1}) for the amplitude of the effective pumping:
\begin{equation}\label{Pk_eff2}
P_k = h_\mathrm{p} V + S (\sum_{\kappa}{n_\kappa}+\sum_{s}{n_s})
e^{i \psi} = h_\mathrm{p} V + S (N_\mathrm{\kappa} + N_\mathrm{s})
e^{i \psi}.
\end{equation}
Here $N_\mathrm{\kappa} = \sum_{\kappa}{n_\kappa} $ and $N_\mathrm{s} = \sum_{s}{n_s}$ are the total
number of magnons per unit of volume for the dominating and standing spin-wave groups, correspondingly,
$S = 2 (\gamma \hbar/M_0) S_{0 0}$ is the renormalized coefficient of pair interaction, and $\psi =
\psi_k$ is the effective phase of the collective magnon pairs.

At a first glance our assumption that the effective phase of all spin-wave groups involved in the
parametric interaction with pumping is approximately the same seems to be rather arbitrary. However, the
theory of parametric interaction of waves \cite{S-theory_Lvov_book} states, that all the excited wave
groups are involved in the renormalization of the effective pumping according to Eqs. (\ref{Pk_eff1}),
(\ref{Pk_eff2}), and that the combined effect of all these groups determines the acting amplitude of the
effective pumping. Thus, as a first approximation, we can assume that the effective phases of different
spin-wave groups do not differ too much.

Another significant simplification of our model is the assumption that all the wave groups participating
in the parametric interaction with pumping are always in exact parametric resonance with it ( i.e.
$\tilde{\omega}_k= \omega_\mathrm{p}/2$)  independently of the effective amplitude of a particular wave
group. In other words, in our model we assume that the nonlinear frequency shift described by the
four-wave nonlinear coefficients $T_{k k_1}$ in Eq.~\ref{omega_eff} does not play a significant role in
the parametric interaction since the wave vectors of the waves participating in this interaction are
automatically adjusted to fulfil the condition of the exact parametric resonance $\tilde{\omega}_k=
\omega_\mathrm{p}/2$ when the  effective amplitude of the wave group changes. We believe that this
simplifying  assumption is reasonable for the conditions of our experiment, because the frequency of
pumping (or the bias magnetic field) in our experiment was always tuned in order to obtain the maximum
amplitude of the restored pulse i.e. the pumping frequency or bias field were always adjusted to achieve
the optimum conditions of parametric interaction.

Using all the above described simplifying assumptions, we can write the equations for the magnon
densities $N_i$  of the two wave groups (where $i=\mathrm{\kappa}$ for the dominating spin-wave group
and $i=\mathrm{s}$ for the standing spin-wave group) and their common phase $\psi$ in the form
\begin{eqnarray}\label{base3}
&&\frac{1}{2}\frac{dN_\mathrm{\kappa}}{dt} = N_\mathrm{\kappa}
[-\Gamma_\mathrm{\kappa} + V h_\mathrm{p} \sin \psi] \nonumber\\
&&\frac{1}{2}\frac{dN_\mathrm{s}}{dt} = N_\mathrm{s}
[-\Gamma_\mathrm{s} + V h_\mathrm{p} \sin \psi]\\
&&\frac{1}{2}\frac{d \psi}{dt} = V h_\mathrm{p} \cos \psi +
 S (N_\mathrm{\kappa} + N_\mathrm{s}) \nonumber,
\end{eqnarray}
where $\Gamma_\mathrm{\kappa}/(2 \pi) = 0.6$~MHz,
$\Gamma_\mathrm{s}/(2 \pi) = 0.69$~MHz are the relaxation parameters
for dominating and standing spin-wave groups, correspondingly. The
physical meaning of the magnon densities $N_\mathrm{\kappa}$ and
$N_\mathrm{s}$ is simple -- they characterize the effective
amplitude of all the spin waves that belong to a corresponding wave
group.

At first, the parametric interaction of both spin-wave groups with pumping leads only to the increase of
the corresponding magnon densities $N_\mathrm{\kappa}$ and $N_\mathrm{s}$. Later, when the the
increasing magnon densities become significant they start to renormalize pumping through the  phase
mechanism described in \cite{S-theory_Lvov_book}, which leads to the limitation of the parametric
amplification for the both spin-wave groups.

Let us now analyze some of the  particular solutions of Eq.~(\ref{base3}). The simplest solution one can
get is the solution for the initial quasi-linear regime, when the system is far from saturation and the
influence of the excited spin wave on the pumping is negligible. It is known \cite{S-theory_Lvov_book}
that the phase $\psi$ in this case is equal $\pi / 2$.  The magnon densities in this quasi-linear regime
can be found as:
\begin{eqnarray}\label{N_linear}
&& N_\mathrm{\kappa} = N_\mathrm{\kappa}^0 \exp{[2 (h_\mathrm{p} V -
\Gamma_\mathrm{\kappa}) t]} \nonumber \\
&& N_\mathrm{s} = N_\mathrm{s}^0 \exp{[2 (h_\mathrm{p} V -
\Gamma_\mathrm{s}) t]},
\end{eqnarray}
where $N_\mathrm{\kappa}^0, N_\mathrm{s}^0$ are the initial magnon densities of the dominating and
standing spin-wave groups. We note that the quasi-linear result (\ref{N_linear}) agrees well with the
previously described qualitative picture of the development of parametric interaction in our
experimental system (see region between points ``\textbf{a}'' and ``\textbf{c}'' in Fig.~\ref{Scheme}).

Another particular solution of the system (\ref{base3}) can be obtained in the stationary regime, when
the pumping is renormalized by the excited spin waves and the system is saturated. It is clear, that in
this stationary regime the magnon densities of both spin-wave groups and their common phase $\psi$ are
constant: $dN_\mathrm{\kappa}/dt = dN_\mathrm{s}/dt = d \psi/dt = 0$. If we assume that in this
saturation regime $\Gamma_\mathrm{s} > \Gamma_\mathrm{\kappa}$ and that the common spin-wave phase
$\psi$ varies in the interval $\pi/2 < \psi < \pi$  we obtain the following stationary solution of the
system (\ref{base3})
\begin{eqnarray}\label{N_stationar}
&& \sin \psi =  \Gamma_\mathrm{\kappa}/(h_\mathrm{p} V) \nonumber \\
&& N_\mathrm{s} = 0\\
&& N_\mathrm{\kappa} = \frac{1}{S} \sqrt{(h_\mathrm{p} V)^2 -
\Gamma_\mathrm{\kappa}^2}. \nonumber
\end{eqnarray}
The analytic result (\ref{N_stationar}) also agrees well with the above presented qualitative picture of
parametric interaction with pumping and it describes the state reached by the system beyond point
``\textbf{d}'' in Fig.~\ref{Scheme}).

To get more detailed information about the temporal behavior of the magnon densities of two magnon
groups we solved the systems of equation (\ref{base3}) numerically, assuming that at the initial moment
the phase $\psi$ is the same as in the quasi-linear case $\psi = \psi^0 = \pi/2$.

The initial value of the magnon density of the dominating spin-wave group is determined by the level of
thermal noise existing at a given temperature in a ferrite film, and can be estimated using standard
methods \cite{S-theory_Lvov_book}. Our estimation performed assuming that the dominating spin-wave group
consists of short-wavelength exchange-dominated spin waves propagating perpendicular to the direction of
the bias magnetic field gave the following value for the total number of magnons per unit volume of
ferrite film
\begin{equation}\label{Nkt}
 N_\mathrm{\kappa T}|_{T=300\mathrm{K}} \approx 10^{12}~1/\mathrm{cm}^3.\nonumber
\end{equation}

This estimation does not take into account the fact that the initial microwave signal could heat-up the
dominating spin-wave group even before the pumping is switched on. However, to account for this effect
we need first to evaluate the number of magnons in the group of standing spin waves created by the input
signal.

The estimation of the initial magnon density for the standing spin-wave group $N_\mathrm{s}^0$ turns out
to be a rather complicated task, because it involves the calculation of the transformation coefficient
of the microwave signal into a propagating MSSW and, also, the coefficient of partial transformation of
the MSSW into the standing (thickness) spin-wave modes of the film. It is clear, however, that the
initial magnon density $N_\mathrm{s}^0$ in the standing spin-wave group must be proportional to the
power of the input microwave signal $N_\mathrm{s}^0 = K_\mathrm{s} P_\mathrm{s0}$. It is also clear that
as soon as the input signal pulse is gone from the film the magnon density of the standing wave group
will decay exponentially with time with the exponent $\Gamma_\mathrm{s}$ equal to the relaxation
parameter of the standing spin waves. Thus, if the delay time between the input signal pulse and the
pumping pulse is $t_\mathrm{p}$ we can evaluate the magnon density in the standing spin-wave group at
the initial moment of parametric amplification (i.e. at the moment when the pumping pule is switched on)
as
\begin{equation}\label{Ns0}
N_\mathrm{s}^0 = K_\mathrm{s} P_\mathrm{s0} \cdot \exp{(-\,2 \,
\Gamma_\mathrm{s} \, t_\mathrm{p})}, \nonumber
\end{equation}
where $P_\mathrm{s0}$ is the power of input signal, $\Gamma_\mathrm{s}/(2 \pi) = 0.69$~MHz is a
relaxation parameter of the standing spin-wave group, $t_\mathrm{p} = 280$~ns is the delay time between
the input sinal pulse and the pumping pulse, and $K_\mathrm{s}$ is a phenomenological coefficient
describing the multi-step transformation of the input microwave signal into the standing spin-wave mode.
For the conditions of our experiment we evaluated the coefficient $K_\mathrm{s}$ as  $K_\mathrm{s} = 4
\cdot 10^{16}$\,$\mathrm{W}^{-1}\mathrm{cm}^{-3}$, which means that for the input signal power
$P_\mathrm{s0}$=10~$\mu$W the initial total magnon density in the standing spin-wave group is equal to
\begin{equation}\label{Ns00}
N_\mathrm{s}^0 = 3.6 \cdot 10^{13}~1/ \mathrm{cm}^3. \nonumber
\end{equation}
Thus, at the initial moment of parametric interaction with pumping the standing spin-wave group in our
case is approximately 40 times stronger than the dominating spin-wave group, caused mostly by thermal
magnons.

In order to make our model more realistic we need to take into account another effect that plays an
important role in the process of microwave signal restoration by parametric pumping. This important
effect is the effect of  elastic two-magnon scattering which is always present in real magnetic systems.
It leads to a partial transfer of energy from one spin-wave groups to another, while the spin-wave
frequency is conserved.

Under the conditions of our experiment the effect of two-magnon scattering  leads to partial transfer of
energy from the standing spin-wave group (excited by the input signal) to the dominating spin-wave
group. Thus, this effect amounts to an effective heating-up of the initial thermal spin wave level. This
heating-up effect was investigated in a previous article \cite{Nonres_Restoration}, where the
nonresonant signal restoration (i.e. the process where the signal frequency was not equal to half of the
pumping frequency) was investigated. Below, we shall use an approach similar to
\cite{Nonres_Restoration}, and we will introduce a phenomenological coefficient $\beta$ which accounts
for this heating-up. In the framework of this phenomenological approach the initial amplitude of the
dominating spin-wave group, after the application of the input signal pulse, could be evaluated as
\begin{equation}\label{Nkapa}
N_\mathrm{\kappa}^0 = N_\mathrm{\kappa T} + \beta N_\mathrm{s}^0,
\nonumber
\end{equation}
where $\beta = 4 \cdot 10^{-2}$\, $\mathrm{cm}^3$ is a
phenomenological parameter describing the efficiency of two-magnon
scattering. Taking into account this two-magnon heating-up effect we
were able to get a more realistic evaluation for the initial magnon
density of the dominating spin-wave group in the case when the input
pulse power was equal to $P_\mathrm{s0}$=10~$\mu$W
\begin{equation}\label{Nkapa0}
N_\mathrm{\kappa}^0 = 2.4 \cdot 10^{12}~1/\mathrm{cm}^3, \nonumber
\end{equation}
which means that the difference between the initial amplitudes of the standing and the dominating wave
groups in our experiment was about 15 times.

Another important feature of our model is the account of nonlinear dissipation in the standing
(signal-generated) spin-wave group. In contrast with the dominating spin-wave group, excited mostly by
the thermal noise and having arbitrary phases of individual spin waves, the standing spin-wave group is
coherent, and all the spin-wave phases in this wave group are determined by the phase of the input
signal. Thus, the spin waves in the standing spin-wave group can effectively participate in four-wave
(second-order) parametric interactions both within the group and with the spin waves belonging to other
groups. This interaction leads to an effective transfer of energy from the standing spin-wave group,
that can be phenomenologically described as nonlinear dissipation by the following form:
\begin{equation}\label{Ns00}
\Gamma_\mathrm{s} = \Gamma_\mathrm{s0} (1 + \eta(N_\mathrm{\kappa} +
N_\mathrm{s})). \nonumber
\end{equation}
The coefficient of nonlinear damping $\eta$ for the conditions of
our experiment was evaluated as  $\eta =
1.7\cdot10^{-20}$\,$\mathrm{cm}^3$.

As it was mentioned in the previous section, the nonlinear damping of spin waves is one of the possible
mechanisms leading to the limitation of parametric amplification. However, we would like to stress one
more time, that in our experiments the nonlinear damping of the standing wave group, described by the
coefficient $\eta$, played a relatively minor role, and the limitation of parametric amplification was
mainly caused by the phase mechanism \cite{S-theory_Lvov_book} described  in our model (\ref{base3}) by
the nonlinear coefficient $S$.

The coefficient $S$ of the nonlinear four-wave interaction between the pairs of excited spin waves was
calculated using the expressions presented in \cite{Magnetization_Oscillations_and_Waves} to give $S
\approx 5 \cdot 10^{-13}$~$\mathrm{cm}^3/\mathrm{s}$.

The pumping magnetic field was calculated from the experimental power of the pumping pulse as $
h_\mathrm{p} V = \sqrt{P_\mathrm{p}} / K_\mathrm{p}$, where $K_\mathrm{p} = 1.07 \cdot 10^{-7}$\,s
$\cdot$ W is the coefficient which describes the efficiency of the open dielectric resonator through
which the pumping pulse was supplied to the ferrite film in our experiment.

The model (\ref{base3}) with the initial conditions and parameters described above was solved
numerically to describe the temporal evolution of the amplitudes of two wave groups participating in the
parametric interaction with pumping.

\section{Results and discussion}

The results of the numerical solution of the system of equations (\ref{base3}) for the above specified
parameters are presented in Fig.~\ref{Waveforms}, Fig.~\ref{idea_graph}, Fig.~\ref{Graph_All} and
Fig.~\ref{input_signal}. We shall discuss below each of these figures.

\subsection{Temporal dynamics of the parametric interaction process }

\begin{figure}
\begin{center}
\scalebox{1}{\includegraphics[width=8.5 cm,clip]{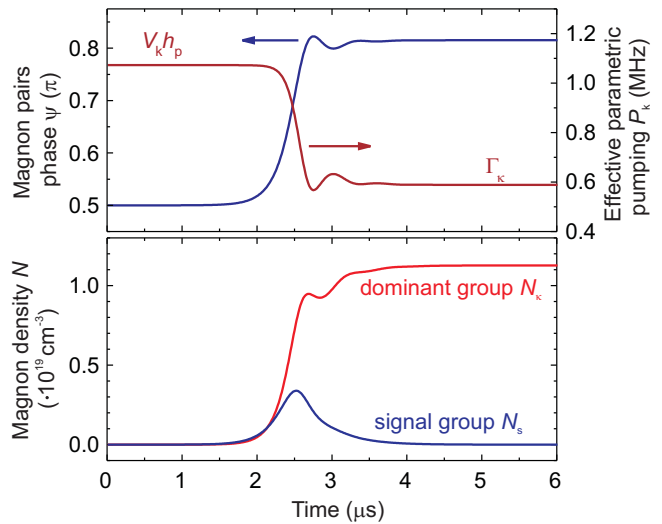}}
\end{center}
\vspace*{-0.7cm}\caption{(color online) Temporal evolution of the
spin-wave effective phase $\psi$ and amplitude of the renormalized
effective pumping (upper frame) and temporal evolution of the magnon
densities $N_\mathrm{\kappa}$ and $N_\mathrm{s}$ of the dominating
and standing spin-wave groups (lower frame). } \label{idea_graph}
\end{figure}

The temporal dynamics of parametric interaction of the two excited spin-wave groups with
constant-amplitude pumping that was switched on abruptly at time $t=0$ is illustrated by
Fig.~\ref{idea_graph}. The upper frame demonstrates the evolution of the collective spin-wave phase
$\psi$ and the amplitude $P_k$ of the renormalized effective parametric pumping, while the lower frame
of Fig.~\ref{idea_graph} illustrates the evolution of the magnon densities $N_\mathrm{\kappa}$ and
$N_\mathrm{s}$ of the  dominating and standing spin-wave groups, respectively.

The numerical results presented in Fig.~\ref{idea_graph} clearly confirm the qualitative picture of the
parametric interaction shown in Fig.~\ref{Scheme}.

In the initial time interval ( $0< t \lesssim 2.2\,\mu$s) a quasi-linear parametric amplification of the
both spin-wave groups takes place. In this quasi-linear regime  the effective spin-wave phase
$\psi=\pi/2$, the effective pumping is not significantly renormalized and is practically equal to $V
h_\mathrm{p}$, while the magnon densities  $N_\mathrm{\kappa}$ and $N_\mathrm{s}$ are very well
described by the analytical result (\ref{N_linear}).

At the final stage of parametric interaction ($t \gtrsim 3.5\,\mu$s), when the system reaches stationary
(or saturation) regime, we also recover the above derived analytic result (\ref{N_stationar}). In this
stationary regime the effective pumping $P_k$ is strongly renormalized by the excited spin waves, and is
stabilized at the value $P_k = \Gamma_\mathrm{\kappa} < V h_\mathrm{p}$, at which it just compensates
the losses of the dominating spin-wave group. Under these conditions the magnon density
$N_\mathrm{\kappa}$ of the dominating spin-wave group becomes constant, while the magnon density
$N_\mathrm{s}$ of the standing spin-wave group vanishes.

In the transitional region ( $2.2\,\mu$s $\lesssim  t \lesssim 3.5\,\mu$s) the amplitude $N_\mathrm{s}$
of the standing spin-wave group first increases, and then begins to decrease, which results in the
appearance of a finite-duration restored pulse at the output of the system.  The theoretically
calculated profiles of the restored output microwave pulse shown in Fig.~\ref{Waveforms}(b) practically
repeat the temporal profile of the magnon density $N_\mathrm{s}(t)$ of the standing spin-wave group
shown in the lower frame of Fig.~\ref{idea_graph}.

\subsection{Restored signal parameters vs pumping power}

\begin{figure}[t]
\begin{center}
\scalebox{1}{\includegraphics[width=8 cm,clip]{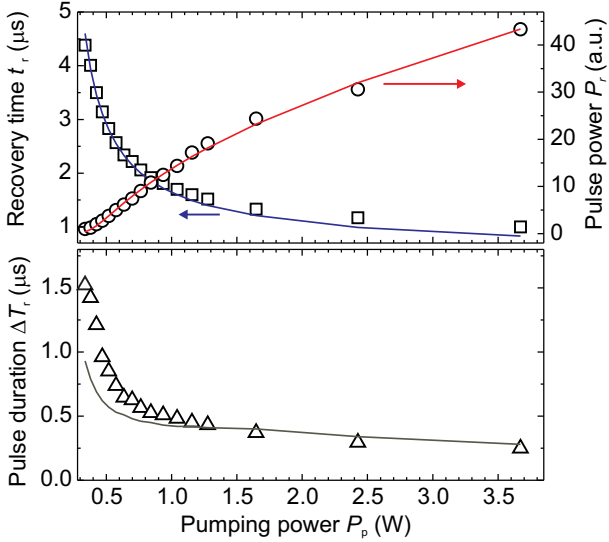}}
\end{center}
\vspace*{-0.5cm}\caption{Delay time (squares), peak power (open
circles), and duration (triangles) of the restored pulse as
functions of the pumping power. Solid lines: results of calculation.
Input signal power $P_\mathrm{s0} = 10 \, \mu$W.} \label{Graph_All}
\end{figure}

As it was mentioned above, the characteristics of the restored microwave pulse, such as the peak power,
duration, and delay time, are mainly determined by the process of parametric interaction with pumping.
This fact is illustrated by Fig.~\ref{Waveforms}(a), where the experimental normalized profiles of the
restored pulses are presented for different values of the pumping power. In Fig.~\ref{Waveforms}(b)
calculated normalized profiles of the restored pulse are presented for the same pumping powers. It is
clear from this figure, that with increasing pumping power both the delay time and duration of the
restored pulse decrease. This effect can be easily explained by the fact that in case of higher pumping
power the system saturates (i.e. reaches a stationary regime) faster, and, as a result, the restored
microwave pulse appears earlier.

The experimental (symbols) and theoretical (solid lines) dependencies of the peak power $P_\mathrm{r}$,
recovery time $t_\mathrm{r}$, and duration $\Delta t_\mathrm{r}$ of the restored pulse on the pumping
power $P_\mathrm{p}$ are shown in Fig.~\ref{Graph_All}. It is evident that the increase of the pumping
power $P_\mathrm{p}$ leads to decrease of $t_\mathrm{r}$ and $\Delta t_\mathrm{r}$, and, at the same
time, to the increase of the peak power $P_\mathrm{r}$ of the restored signal. The above shown behavior
of the restored pulse peak power can be explained by that fact that with increasing pumping power the
amplification rates $\exp[( h_\mathrm{p} V-\Gamma_\mathrm{\kappa}) t]$ and  $\exp[( h_\mathrm{p}
V-\Gamma_\mathrm{s} ) t]$ become closer and the difference in the peak amplitudes of the standing and
dominating spin-wave groups becomes smaller. Thus, as a result, the amplitude of the restored pulse
becomes larger.

It is clear from Fig.~\ref{Graph_All} that the theoretical results for the restoration time and power of
the restored pulse are in good agreement with experiment. At the same time, the calculated dependence of
the restored pulse duration on the pumping power agrees with the experiment only qualitatively. We
believe that a significantly  better description of the experimental results could be achieved if the
theoretical model is modified to take into account such experimental parameters as the finite duration
of the input signal, realistic widths of the dipole-exchange gaps in the ferrite film spectrum, and the
frequency selective character of the parametric amplification process.

\subsection{Restored signal parameters vs input signal power}

Using our theoretical model we also calculated the influence of the input signal power $P_\mathrm{s0}$
on the parameters of the restored microwave pulse.

According to the simple analytical model presented in \cite{PRL_restoration} where the increase of the
input signal power the power of the restored pulse increases linearly, while the recovery time should
remain constant.

\begin{figure}[t]
\begin{center}
\scalebox{1}{\includegraphics[width=8.5 cm,clip]{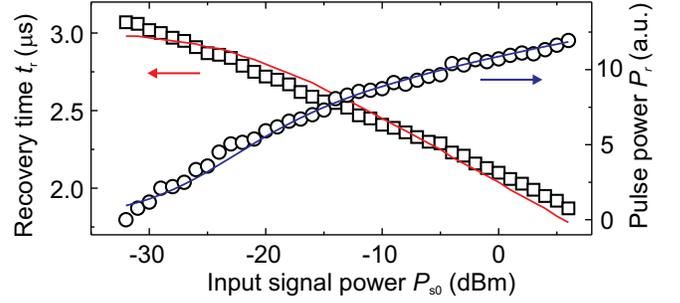}}
\end{center}
\vspace*{-0.5cm}\caption{Delay time (squares) and peak power (open
circles) of the restored pulse as functions of the input signal
power. Solid lines: results of calculation. Pumping power
$P_\mathrm{p} = 0.52$\,W.} \label{input_signal}
\end{figure}

At the same time, our experiments (see Fig.~\ref{input_signal}) clearly demonstrate that the increase of
the input signal power $P_\mathrm{s0}$ leads to decrease of the recovery time $t_\mathrm{r}$ and to
increase of the restored pulse power $P_\mathrm{r}$, with the experimentally observed increase being not
linear.

It is evident from Fig.~\ref{input_signal} that our new theoretical model (\ref{base3}) describes the
experiment rather well and, therefore, significantly improves the simple analytical model presented in
\cite{PRL_restoration}. There are two reasons why our new model (\ref{base3}) improves the accuracy of
the theoretical description. First of all, in the analytical model \cite{PRL_restoration} only the
dominating spin-wave group participates in the process of saturation of the  parametric amplification,
while in reality both spin-wave groups significantly influence the saturation process. Another reason,
which also plays an important role in the parametric interaction process, is the heating of the thermal
magnons by the input signal (see \cite{Nonres_Restoration} for details). This effect was taken into
account in the model by the introduction of the phenomenological coefficient $\beta$.

\section{Summary and conclusion}

In conclusion, we investigated the effect of storage and recovery of a microwave signal in a ferrite
film having discrete dipole-exchange spin-wave spectrum. The signal is stored in the form of standing
spin waves (thickness modes) excited by the input microwave pulse in the frequency intervals near the
dipole-exchange gaps in the spin-wave spectrum of a ferrite film. The signal is recovered in the process
of parametric interaction of these standing spin-wave modes forming a trail of the input pulse with a
long and powerful pumping pulse having the carrier frequency that is close to the double frequency of
one of the dipole-exchange gaps in the ferrite film spectrum. An approximate (mean field) theoretical
model, taking into account the competition of two spin-wave groups interacting with parametric pumping,
and involving a nonlinear phase mechanism of limitation of parametrical amplification was used to
describe the experimental results. The developed model provides a good quantitative description of the
experimentally observed effect of storage and recovery of microwave pulses in ferrite films.

This work was supported by the Deutsche Forschungsgemeinschaft (SFB/TRR~49), by the Ukrainian Fund for
Fundamental Research (25.2/009), by the MURI grant W911NF-04-1-0247 from the U.S. Army Research Office,
and by the Oakland University Foundation.

\end{document}